\documentclass[aps,prl,reprint,showpacs,showkeys,preprintnumbers,superscriptaddress]{revtex4-2}

\usepackage[bookmarks=true]{hyperref}
\usepackage{graphicx,epsfig,color}
\usepackage[english]{babel}
\usepackage{amssymb,amsfonts,amsmath}
\usepackage{verbatim}
\usepackage{float}

\newcommand{\be}{\begin{equation}} \newcommand{\ee}{\end{equation}}
\newcommand{\bea}{\begin{eqnarray}} \newcommand{\eea}{\end{eqnarray}}

\def\al{\alpha} 
\newcommand{\ben}{\begin{equation}}
\newcommand{\een}{\end{equation}}
\newcommand{\ba}{\begin{array}}
\newcommand{\ea}{\end{array}}
\newcommand{\bit}{\begin{itemize}}
\newcommand{\eit}{\end{itemize}}

\newcommand\AdS[1]{AdS$_{#1}$}

\newcommand{\HN}{H_\text{n}}

\newcommand{\Mpl}{M_{\text{P}}}

\newcommand\rhor{r_H}

\newcommand{\TN}{T_\text{n}} % Nucleation temperature
\newcommand{\Tn}{T_\text{n}} % Nucleation temperature
\newcommand{\Tc}{T_\text{c}} % Critical temperature
\newcommand{\vw}{v_\text{w}} % Wall speed

\newcommand{\inlinesection}[1]{\textit{#1}.-}

\begin{document}

\title{Gravitational Waves at Strong Coupling from an Effective Action}

\author{F\"eanor Reuben Ares}
\email{F.R.Ares@sussex.ac.uk}
\affiliation{Department of Physics and Helsinki Institute of Physics\\
P.O.~Box 64, FI-00014 University of Helsinki, Finland}
\affiliation{Department of Physics \& Astronomy, University of Sussex\\
Brighton, BN1 9QH, United Kingdom}

\author{Oscar Henriksson}
\email{oscar.henriksson@helsinki.fi}
\affiliation{Department of Physics and Helsinki Institute of Physics\\
P.O.~Box 64, FI-00014 University of Helsinki, Finland}

\author{Mark Hindmarsh}
\email{mark.hindmarsh@helsinki.fi}
\affiliation{Department of Physics and Helsinki Institute of Physics\\
P.O.~Box 64, FI-00014 University of Helsinki, Finland}
\affiliation{Department of Physics \& Astronomy, University of Sussex\\
Brighton, BN1 9QH, United Kingdom}

\author{Carlos Hoyos}
\email{hoyoscarlos@uniovi.es}
\affiliation{Department of Physics, Universidad de Oviedo and Instituto de Ciencias y Tecnolog\'{\i}as Espaciales de Asturias (ICTEA)\\
c/ Federico Garc\'{\i}a Lorca 18, ES-33007 Oviedo, Spain}

\author{Niko Jokela}
\email{niko.jokela@helsinki.fi}
\affiliation{Department of Physics and Helsinki Institute of Physics\\
P.O.~Box 64, FI-00014 University of Helsinki, Finland}

\begin{abstract}
Using a holographic derivation of a quantum effective action for a scalar operator at strong coupling, we compute quasi-equilibrium parameters relevant for the gravitational wave signal from a first order phase transition in a simple dual model. We discuss how the parameters of the phase transition vary with the effective number of degrees of freedom of the dual field theory. Our model can produce an observable signal at LISA if the critical temperature is around a TeV, in a parameter region where the field theory has an approximate conformal symmetry.
\end{abstract}

\preprint{HIP-2021-24/TH}


\maketitle

%%%%%%%%%%%%%%%%%%%%%%%%%%%%
%%%%%%%%%%%%%%%%%%%%%%%%%%%%
\inlinesection{Introduction}
%%%%%%%%%%%%%%%%%%%%%%%%%%%%
%%%%%%%%%%%%%%%%%%%%%%%%%%%%
A first order phase transition in the early Universe \cite{Coleman:1977py,Linde:1978px,Steinhardt:1981ct,Linde:1981zj} would generate gravitational waves (GWs) \cite{Witten:1984rs,Hogan:1986qda}.  If the critical temperature of the transition were around the electroweak scale 0.1 -- 1 TeV, the GWs would be potentially observable at future space-based detectors, such as the Laser Interferometer Space Antenna (LISA) \cite{Audley:2017drz,Caprini:2019egz}, while a critical temperature around the scale of confinement of the strong interaction (100 MeV) is of interest for pulsar timing arrays. Recent reports of a possible signal at NANOgrav \cite{NANOGrav:2020bcs}, which if confirmed would likely be from merging supermassive black holes \cite{Middleton:2020asl}, have also prompted an examination of phase transitions as a source \cite{NANOGrav:2021flc}.

In the Standard Model it is well established that both the confinement and electroweak transitions are crossovers \cite{Borsanyi:2016ksw,Kajantie:1996mn,Laine:1998vn,Laine:2012jy}. However, the Standard Model is incomplete: for example, it does not account for the dark matter in the Universe or the baryon asymmetry (see e.g.~\cite{Cline:2018fuq} for a pedagogical review). Numerous extensions have been put forward to solve these and other problems, which would also induce a first order electroweak transition (see {e.g.} \cite{Weir:2017wfa,Caprini:2019egz} for reviews). Hence a search for GWs from the early Universe is also a search for physics beyond the Standard Model. 

A first order phase transition in the early Universe would proceed through the nucleation, expansion and merger of bubbles of the stable phase \cite{Guth:1981uk,Steinhardt:1981ct,Enqvist:1991xw,Turner:1992tz}, (see \cite{Laine:2016hma,Hindmarsh:2020hop} for pedagogical reviews). The consequent disturbances in the cosmic fluid would produce GWs \cite{Witten:1984rs,Hogan:1986qda}. Much progress has been made recently towards an accurate understanding of the process \cite{Caprini:2019egz}, with the aim of enabling LISA to probe the physics of an era that is difficult to explore otherwise.

However, if the phase transition occurs at strong coupling, we are confronted by the difficulty of computing thermodynamic and transport properties. In this letter, we present a consistent strong-coupling framework for the calculation of the quasi-equilibrium properties most relevant for GW production, and illustrate its use with a simple model. 

The GW signal from a first order phase transition depends on four main parameters: the nucleation temperature $\Tn$, the transition rate $\beta$, the dimensionless transition strength parameter $\alpha$, and the wall (phase boundary) speed $\vw$. The speed of sound also affects the signal \cite{Giese:2020rtr,Giese:2020znk}. The critical temperature of the phase transition $\Tc$ sets the scale. These parameters control the conversion of energy into fluid motion and are directly connected to the detailed shape of the GW power spectrum \cite{Hindmarsh:2017gnf,Hindmarsh:2019phv}, through which they are accessible at LISA \cite{Gowling:2021gcy}. Hence their calculation is of utmost importance to the drive to use GW detectors to probe high energy physics. 

At weak coupling perturbative methods can give good results for the quasi-equilibrium parameters $\Tn$, $\beta$, and $\al$ (for recent discussion of the calculations and their uncertainties see \cite{Gould:2019qek,Croon:2020cgk,Gould:2021oba}). 
In general, $\vw$ is a fully non-equilibrium quantity that has been computed only in various approximations \cite{Liu:1992tn,Moore:1995si,John:2000is,Huber:2011aa,Bodeker:2017cim,Dorsch:2018pat,aleks2020bubble,Friedlander:2020tnq}. If, however, the extension to the Standard Model is a strongly coupled field theory the parameters are much more difficult to calculate.  Historically, lattice methods have been used for the strictly equilibrium quantities in specific theories, the critical temperature and the latent heat: for example, it is known that SU($N$) Yang-Mills theory, where $N$ is the number of colours or independent charges, has a first order confinement transition for $N\ge3$ (see e.g. \cite{Lucini:2012gg}). GW production in such theories has been studied in \cite{Huang:2021aou,Halverson:2020xpg}. The functional renormalisation group has recently been used for GW production in a scalar field theory at strong coupling \cite{Croon:2021vtc}.

In recent years, holography has proved a powerful tool to rework the problem, equating field theories with string theories in a larger number of dimensions \cite{Maldacena:1997re,Witten:1998qj}.  Quantities in a field theory with a large number of degrees of freedom at strong coupling are computable from classical solutions in the string theory, which are essentially solutions to Einstein equations with various fields as sources of energy-momentum.
Using holography, thermodynamic properties of phase transitions have been studied in so-called ``bottom-up'' models (where the source fields are not formally derived from a string theory) \cite{Attems:2016ugt,Gursoy:2018umf,Bea:2018whf}, and GWs have been considered in the context of neutron star mergers \cite{Jokela:2018ers,Ecker:2019xrw,Jokela:2020piw} and phase transitions in the early Universe \cite{Ares:2020lbt}. Recently there has also been progress in finding the wall speed \cite{Bea:2021zsu,Bigazzi:2021fmq,Henriksson:2021zei}.

In this letter, we outline a new method for calculating the quasi-equilibrium parameters $\al$, $\beta$, and $\Tn/\Tc$.  The method uses a quantum effective action, which we show that it can be derived using holography, giving full details in \cite{Ares:2021ntv}. With it we construct bubble solutions taking the system to the stable phase directly in the field theory, avoiding the need to solve partial differentials in the gravity dual. The computed quantities are then used to determine the corresponding signals using current models of GW production \cite{Caprini:2019egz}. The scaling of the results with $N$ in the putative gauge theory is discussed and scans for all quantities are shown for $N = 8$, where the holographic assumption of large $N$ should still be valid. Here we define $N$ from $L^3/\kappa_5^2 = N^2$, where $\kappa_5^2$ is the 5D gravitational constant and $L$ the radius of curvature.

We find that the large $N$ restriction generically pushes $\beta/\HN$ (where $\HN$ is the nucleation Hubble rate) to high values; $10^3 - 10^8$ in this particular model for $N = 8$, with the vast majority of values above $10^5$. This restricts a detectable GW signal to a corner of parameter space where the minima in the effective potential are far apart and breaking of conformal invariance in the trivial vacuum is $1/N$ suppressed.  In this region, a phase transition with critical temperature around 1 TeV would be observable, which is around the scale where one would expect physics beyond the Standard Model to appear.

%%%%%%%%%%%%%%%%%%%%%%%%%%%%
%%%%%%%%%%%%%%%%%%%%%%%%%%%%
\inlinesection{Effective action from holography}
%%%%%%%%%%%%%%%%%%%%%%%%%%%%
%%%%%%%%%%%%%%%%%%%%%%%%%%%%
We start with a free scalar field $\phi$ in five dimensions with action
\be\label{eq:bulkaction}
 S_{\textrm{bulk}} = \frac{1}{2\kappa_5^2}\int d^5 x \sqrt{g} \left(\mathcal{R}+\frac{12}{L^2}-\left(\partial\phi\right)^2- m^2\phi^2\right) \
\ee
where $\mathcal{R}$ is the Ricci scalar and $m$ the mass parameter. We will set $L = 1$ hereafter. We are interested in homogeneous, isotropic solutions that are asymptotically \AdS{5} with a black brane in the interior; a suitable ansatz is
\be\label{eq:metric}
 ds^2 = -e^{-2\chi(r)}h(r) dt^2 + \frac{dr^2}{h(r)}+r^2d\vec x^2 \ , \ \ \phi=\phi(r) \ .
\ee
Such a black brane solution is dual to a field theory state with temperature $T={e^{-\chi(\rhor)}h'(\rhor)}/{4\pi}$ and entropy density $s={2\pi\rhor^3}/{\kappa_5^2}$, both evaluated at the horizon radius $\rhor$ of the black brane, where $h(\rhor) = 0$. Fixing $T$, one finds a one-parameter family of solutions. At the boundary $r\to\infty$, the scalar field falls off as $\phi\sim \phi_-/r^{\Delta_-}+\phi_+/r^{\Delta_+}$, where $\Delta_\pm= 2\pm\sqrt{4+m^2}$. The one-parameter family of solutions determines $\phi_+$ as a function of $\phi_-$; this can be related to the generating functional of a conformal field theory (CFT) in Minkowski space, defined on the boundary $r \to \infty$.

We will use here ``alternative quantisation" in which $\phi_+$ determines the source of a field operator $\Psi$ of the CFT, and $\phi_-$ is related to the expectation value $\langle\Psi\rangle$ \cite{Klebanov:1999tb}. 
Choosing this quantisation allows us to deform the CFT by the operators $\Psi$, $\Psi^2$, and $\Psi^3$, with couplings $\Lambda$, $f$, and $g$, respectively. The deformations, which are implemented through the choice of boundary conditions at $r \to \infty$ \cite{Witten:2001ua}, result in a theory with first order thermal phase transitions for suitable parameters.
We take the cubic term to be exactly marginal (scaling dimension 4) which amounts to choosing $m^2 = -{32}/{9}$ in (\ref{eq:bulkaction}). Thus the scaling dimensions for $\Lambda$ and $f$ are $8/3$ and $4/3$, respectively. 

We therefore have three scales $T$, $\Lambda$, and $f$ which are assembled into two dimensionless ratios, chosen to be $\Lambda_f = \Lambda/f^{2}$ and $\widetilde{T} = T/(|\Lambda|^{3/8}+|f|^{3/4})$. The overall scale is a free parameter at this simplified level. 

The boundary field theory effective action at $T$ is defined as a functional of field expectation value $\psi$ through 
\be
 \Gamma_T[\psi] = W_T[J] - N^2 \int d^4 x\, \psi J \ ,
\ee
with $W_T[J]$ being the generating functional in the presence of a source $J$, and the factor of $N^2$ appearing due to the definition $\psi=W_T'[J]/N^2$. For static configurations, the first two terms in the derivative expansion are 
\be\label{eq:effectiveaction}
\Gamma_T[\psi] = - N^2\int d^4 x\left(V_T(\psi)+\frac12 Z_T(\psi)(\nabla \psi)^2 \right) \ ,
\ee
where $V_T(\psi)$ is the effective potential. By using the holographic equivalence of the renormalised on-shell gravitational action with the generating functional \cite{Hertog:2004ns,Faulkner:2010gj,Kiritsis:2012ma}, and assuming homogeneous solutions, one can find the effective potential \cite{Ares:2021ntv}, giving 
\be\label{eq:effectivepotential}
 V_T(\psi) = \frac{h_2(\psi,T)}{2} + \frac{7}{9}\psi\,\phi_+(\psi,T) + \Lambda\psi + \frac{f}{2} \psi^2 + \frac{g}{3} \psi^3 \ .
\ee
Here $h_2$ comes from the boundary fall-off of the metric function $h \sim r^2+4\phi_-^2/9r^{2/3}+h_2/r^2$, and $\psi = -\frac43\phi_-$.

To extract the coefficient of the kinetic term $Z_T(\psi)$ we note that the full quadratic part of $\Gamma_T[\psi]$ equals the inverse of the two-point function of $\Psi$. In momentum space, $Z_T(\psi)$ is then given by the coefficient of the $k^2$ term in a low-momentum expansion of the inverse of the two-point function. On the holographic side this can be computed by a standard fluctuation analysis \cite{Kovtun:2005ev}. For our solutions, the $k^4$ term is negligible \cite{Ares:2021ntv}, validating the derivative expansion.

Fixing the theory means fixing $\Lambda_f$ and $g$; here we restrict to the region $-\infty<\Lambda_f\le0$ and $0\le g<\gamma_3\approx 0.278$ ($g>\gamma_3$ renders the potential unbounded from below). In a large part of this, shown in colour in the figures below, the theory displays a first order thermal phase transition.

%%%%%%%%%%%%%%%%%%%%%%%%%%%%%%%%%%%%%%%%%%%%%%%%%%%%%%
\begin{figure*}[ht!]
\center
      \includegraphics[width=0.45\textwidth]{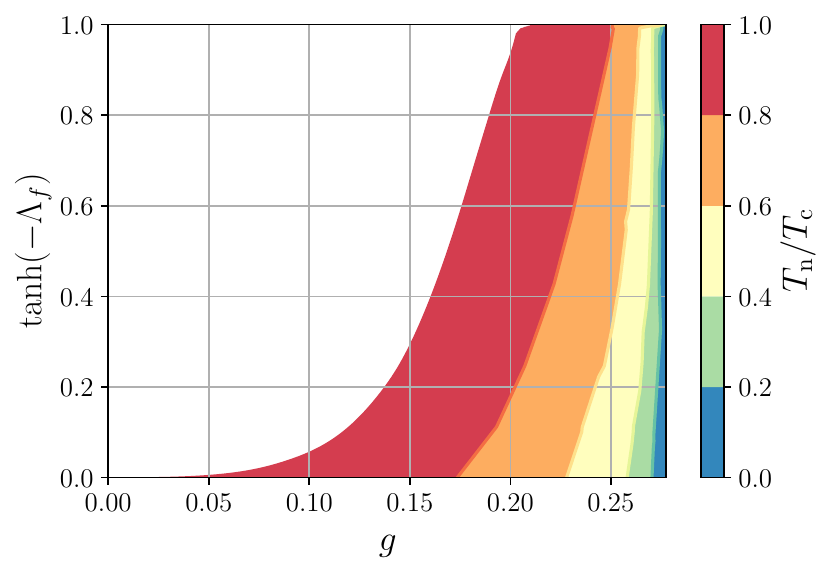}
$\;\;\;\;\;$ 
	\includegraphics[width=0.46\textwidth]{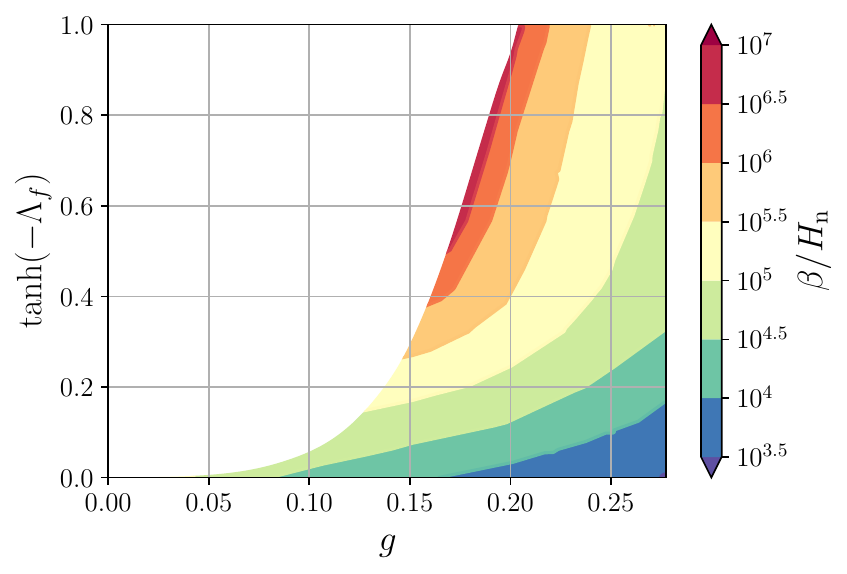}
\caption{Scans of the nucleation temperature $T_\mathrm{n}/T_\mathrm{c}$ (left) and the transition rate $\beta/H_\mathrm{n}$ at $\Tn$ (right).} 
 \label{fig:Tnbeta}
\end{figure*}
%%%%%%%%%%%%%%%%%%%%%%%%%%%%%%%%%%%%%%%%%%%%%%%%%%%%%%
%

%%%%%%%%%%%%%%%%%%%%%%%%%%%%
%%%%%%%%%%%%%%%%%%%%%%%%%%%%
\inlinesection{Gravitational wave parameters}
%%%%%%%%%%%%%%%%%%%%%%%%%%%%
%%%%%%%%%%%%%%%%%%%%%%%%%%%%
%
We can use the flat-space field theory we have constructed to study phase transitions in the early Universe, as relaxation rates at temperature $T$ are expected to be much faster than the Hubble rate $H(T)$.  The phase transition proceeds through localised fluctuations of $\psi$ into the stable phase, just large enough so that the pressure difference overcomes the surface tension. The most probable fluctuation, the critical bubble, is in the form of a bubble with a spatial O$(3)$ symmetry, invariant in the periodic imaginary time coordinate \cite{Linde:1981zj}. The rate per unit volume of bubble nucleation $p(t)$ increases rapidly from zero below $\Tc$, a change quantified by the transition rate parameter $\beta = -d\log(p)/dt$. To a good approximation it can be written $p(t) = p_0 \exp(-\Gamma_b(T))$, where $\Gamma_b$ is the Euclidean action for the critical bubble, whose time dependence is a consequence of the non-zero cooling rate in the expanding Universe. The transition rate parameter is evaluated at $\TN$, the peak of the globally-averaged bubble nucleation rate per unit volume. Hence, given that the temperature decreases as $dT/dt = -H(T)T$, 
\be\label{B_H}
 \beta/\HN = T\frac{d}{dT}\Gamma_b(T) \Big|_{T_\mathrm{n}} \ .
\ee
To find the critical bubble, we extremise the O$(3)$-symmetric action 
\begin{equation}
 \Gamma_{\textrm{O}(3)} = \frac{4\pi N^2}{T}\int d\rho\, \rho^2\left(\frac12Z_T(\psi)\left(\psi' \right)^2+V_T(\psi)\right) \ ,
\end{equation}
looking for solutions representing a bubble of stable phase surrounded by metastable phase. We solve numerically the resulting Euler-Lagrange equation with boundary conditions $\psi(\infty)=0=\psi'(0)$, where the field is defined to vanish at the metastable minimum, and $\psi(0)$ is the shooting parameter. The asymptotic boundary condition is imposed at a suitably large finite radius, which we take to be $20(|\Lambda|^{3/8}+|f|^{3/4})$.

%%%%%%%%%%%%%%%%%%%%%%%%%%%%%%%%%%%%%%%%%%%%%%%%%%%%%%
\begin{figure*}[ht!]
\center
      \includegraphics[width=0.47\textwidth]{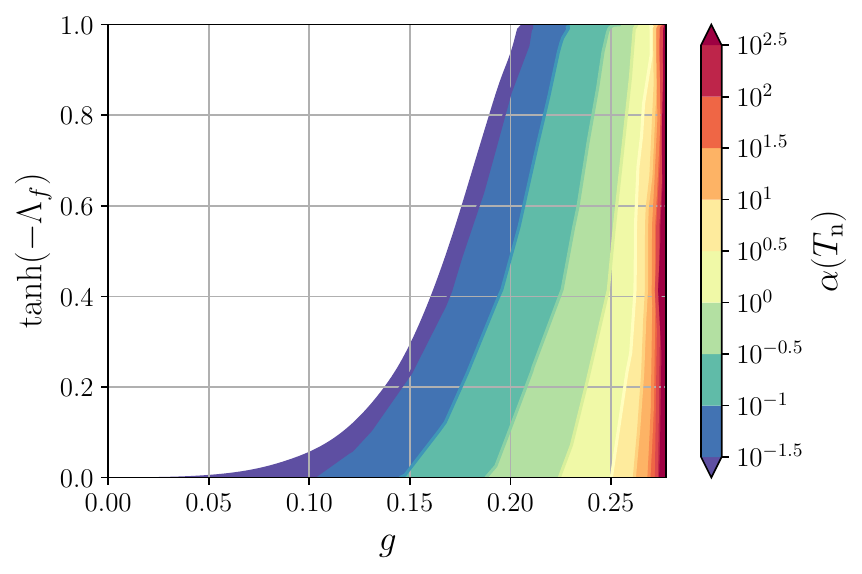}
$\;\;\;\;\;$ 
      \includegraphics[width=0.46\textwidth]{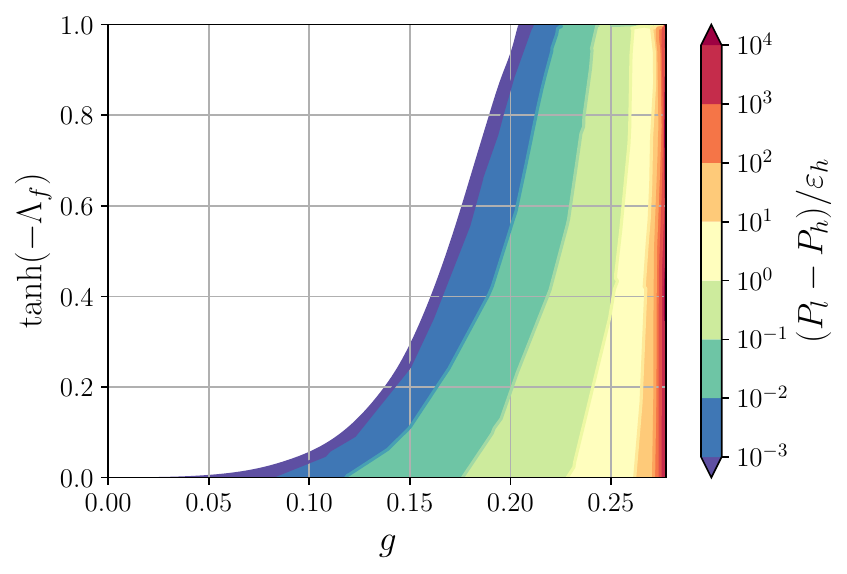}
  \caption{Scan of the transition strength $\alpha$ (left) and pressure change over energy density at $\Tn$ (right).}
  \label{fig:strengthandspeed}
\end{figure*}
%%%%%%%%%%%%%%%%%%%%%%%%%%%%%%%%%%%%%%%%%%%%%%%%%%%%%%

The phase transition can be thought to start when the nucleation rate per unit volume reaches one bubble per Hubble volume per Hubble time, that is, $p = H^4$. The nucleation temperature is reached shortly after, so an approximation to $\Tn$ can be found through $\Gamma_{b}(\Tn) \sim 4\:\log\left({\Mpl}/{\Tc}\right)$. Hence, for $T_\text{c}\approx$ 100 GeV,  bubble nucleation occurs when the action drops to about 150 \cite{Enqvist:1991xw}. 

To understand how the results depend on $N$, note that the bubble action $\Gamma_b$ is generally a monotonic function of temperature below $\Tc$. The action diverges quadratically \cite{Enqvist:1991xw} at $\Tc$ and goes to zero at some lower temperature $T_0$ where the effective potential barrier between the vacua vanishes. As the pre-factor of the action scales as $N^2$, sufficiently large $N$ will push $\TN$ down towards $T_0$. We call this the large supercooling case. We assume that the temperature dependence near $T_0$ is a power law $\Gamma_b \sim N^2(T - T_0)^x$ with $x>0$, the form followed by theories with a canonical kinetic term and a quartic potential,  where $x = 3/2$ \cite{Enqvist:1991xw}.  Fitting a similar power law to our data, we find a value of $x \approx 1.4 - 1.5$. Eq.~(\ref{B_H}) and the definition of $T_\mathrm{n}$ then quickly lead to $\beta/H_\mathrm{n} \sim N^{2/x}$. Thus, for large $N$, $\beta/H_\mathrm{n}$ increases with $N$.

In practice, we are interested in finite but large $N$. Then, it is possible that instead $\TN \approx \Tc$. In this small supercooling case, one can approximate the solution as a so-called ``thin wall'' bubble, consisting of a large ball of the stable phase surrounded by a spherical phase boundary, thin compared with its radius. In this case $\Gamma_b \sim N^2(\Tc - T)^{-2}$,\footnote{Note that our considerations imply that the surface tension of the phase boundary is proportional to $N^2$, consistent with lattice results in SU($N$) gauge theories \cite{Lucini:2003zr,Lucini:2005vg,Lucini:2012gg}. However, lattice results also permit models with a different $N$-dependence \cite{Halverson:2020xpg,Huang:2020crf}.} which leads to $\beta/\HN \sim N^{-1}$, decreasing with $N$.  Thus there can exist models with an ``optimal'' value of $N$ which minimises $\beta/\HN$ while still being large enough for the large-$N$ limit to give accurate results at leading order. In fact, for certain parameter values this is the case for our holographic model; however, despite this the $\beta/\HN$ values remain large. The full range of $\beta/\HN$ for our parameter space is displayed in Fig.~\ref{fig:Tnbeta} on the right, along with the ratio $\TN/\Tc$ in the left plot. The small supercooling limit  $T_\text{n} \approx T_\text{c}$ is approached at the left-most boundary for both plots. 

The energy available for conversion into fluid motion is quantified by the transition strength $\alpha$, which depends on the enthalpy density $w = Ts$ and the pressure $P$ in the two phases. Writing $\theta = w/4 - P$, the transition strength parameter is then defined as \cite{Espinosa:2010hh,Hindmarsh:2019phv} 
\ben
\label{e:AlpDef}
\alpha = \left.\frac{4}{3}\frac{\theta_h(T)-\theta_l(T)}{w_h(T)} \right|_{\TN}\ , 
\een
where subscripts $h$ and $l$ denote the phases stable at high and low temperatures, respectively.

The enthalpy density can be found from the solution to the gravity dual, $\kappa_5^2 Ts = -2h_2-\frac{16}{9}\phi_+\psi $, and the pressure is available from $V_T$ evaluated at its minima. The values for $\alpha$ are shown in the left plot of Fig.~\ref{fig:strengthandspeed}.

The $N$ dependence of $\alpha$ in cases of small and large supercooling follows from linear expansion of $\al$ near a reference temperature $T_*$, $\alpha(\TN) = \alpha(T_*) + \alpha'(T_*)( \Tn - T_*)$, where $T_*$ is either $\Tc$ or $T_0$.  The values $\alpha(T_*)$, being ratios, are independent of $N$.  However, the next term grows with $N$ in the small supercooling case, and decreases as $N^{-2/x}$ in the large supercooling case.

We do not yet have a simple way to calculate the bubble wall velocity  $v_\textrm{w}$. To estimate the wall speed, we adapt a result from \cite{Bea:2021zsu,Bigazzi:2021fmq,Henriksson:2021zei} that at small velocities, $v_\textrm{w}$ is proportional to the pressure difference divided by the high-$T$ phase energy density at $\TN$. To extrapolate to larger velocities, we assume 
\ben
u_\text{w} = \gamma_\text{w} v_\text{w} = C \left. \frac{P_l-P_h}{\varepsilon_h} \right|_{\TN} ,
\een
where $C$ is an O(1) constant and $\gamma_\text{w}$ is the Lorentz factor.  The pressure difference divided by the energy density is shown in the right plot of Fig.~\ref{fig:strengthandspeed}; to estimate the wall speed we set $C=1$. It is not important to get a precise value for $u_\text{w}$ at high $\gamma_\text{w}$, as the hydrodynamic solution for the flow set up by the expanding bubble, and hence the GW signal, depends only on $v_\text{w}$. The same argument for $N$ scaling can be made for $u_\text{w}$ as can be made for $\alpha$.

\begin{figure}[ht!]
\center
      \includegraphics[width=0.49\textwidth]{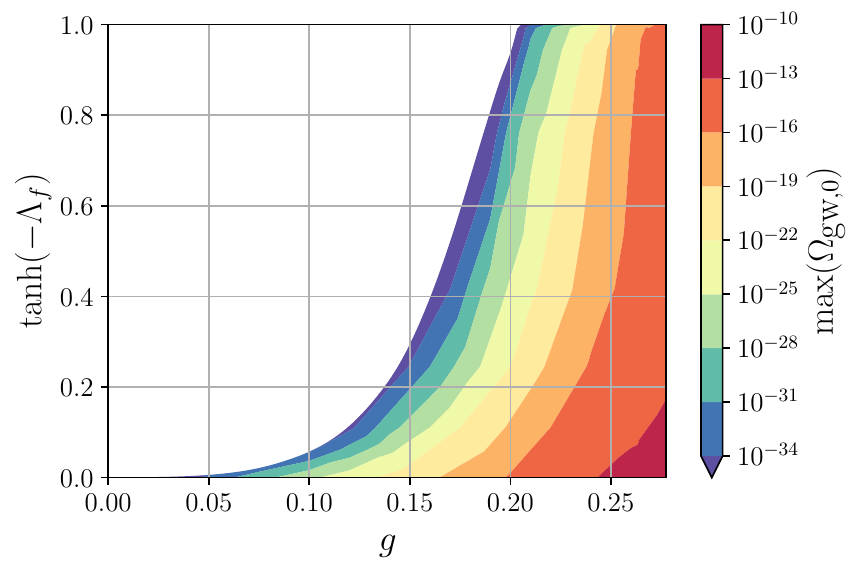}
  \caption{Maximum GW power spectrum, combining Fig.~\ref{fig:Tnbeta} and \ref{fig:strengthandspeed} data, using the model of \cite{Caprini:2019egz} and \cite{Ares:2020lbt}.
  }
  \label{fig:spectra}
\end{figure}

Finally, collating the information gained on $\alpha$, $\beta/H_\mathrm{n}$, $T_\mathrm{n}$, and $v_\textrm{w}$ we calculate the maximum of the GW power spectrum $\Omega_{\textrm{gw},0}$, and the frequency at which it occurs in units of $T_\text{c}$. We use the standard LISA Cosmology Working Group model \cite{Caprini:2019egz}, improved with a numerical kinetic energy suppression factor \cite{Cutting:2019zws}, as described in \cite{Ares:2020lbt}. We take $c_s^2=1/3$, as in the region where there is strong supercooling (and a detectable signal) the sound speed is close to the conformal value. We plot $\max(\Omega_{\textrm{gw},0})$ as a function of our parameters in Fig.~\ref{fig:spectra}.

The maximum of the spectrum, which is independent of the temperature of the phase transitions, takes a broad range of values between $10^{-34}$ and $10^{-10}$). A value above about $10^{-13}$ would be observable at LISA, if the peak frequency was in the range of highest sensitivity $10^{-2}-10^{-3}$ Hz. We find that $\Tc$ would need to be in the range 0.3 to 1.8 TeV for a signal to be detected. This puts the critical temperature in a range relevant for models of strong dynamics leading to electroweak symmetry breaking, such as composite Higgs models (see \emph{e.g.} \cite{Bellazzini:2014yua} for a review).

%%%%%%%%%%%%%%%%%%%%%%%%%%%%
%%%%%%%%%%%%%%%%%%%%%%%%%%%%
\inlinesection{Discussion}
%%%%%%%%%%%%%%%%%%%%%%%%%%%%
%%%%%%%%%%%%%%%%%%%%%%%%%%%%
In this letter we outlined the construction of the effective action for a holographic strongly coupled field theory, and used it to compute the equilibrium and quasi-equilibrium quantities relevant for GW production in a first order phase transition in the early Universe. Details of the construction of this action are presented in \cite{Ares:2021ntv}.  The effective action describes a scalar field at non-zero temperature, computed in a derivative expansion.  That such an action is needed to describe a phase transition has already been argued \cite{Ghosh:2021lua,Janik:2021jbq}; it is also known that hydrodynamics alone is insufficient to describe the bubble's evolution after nucleation \cite{Bea:2021zsu}. 

We illustrated the effective action method with a simple holographic 5D theory with a massive free scalar, which in alternative quantisation is dual to a 4D CFT that can be deformed by simple relevant or marginal operators.  The theory has first order transitions over a wide region of dimensionless coupling ratio space. 
 
Using an estimate for the phase boundary speed motivated by numerical simulations of a similar system \cite{Bea:2021zsu}, we  computed the GW power according to current state of the art \cite{Caprini:2019egz,Hindmarsh:2019phv,Ares:2020lbt}. While the transition is supercooled and strong over a large parameter region, in the sense that a large fraction of the available potential energy is converted into kinetic energy of the fluid, the transition is also generally rapid, completing in less than $10^{-3}$ of the Hubble time, which reduces the signal strength.  In our parameterisation of the model only a relatively small region would be observable at LISA, if the critical temperature is around 1 TeV.  The favoured region has relatively small coupling $\Lambda\approx 0$ and a cubic coupling $g$ close to the boundedness limit.

In the parameter range leading to an observable signal, the phenomenology of the holographic model conforms quite well with the nearly-conformal dynamics described in \cite{Konstandin:2011dr}, including large supercooling followed by a strong transition and a peaked frequency in the millihertz range with a critical temperature of the order of TeV. The nearly-conformal physics can be understood from the fact that when $\Lambda=0$, the breaking of conformal invariance by the coupling $f$ in the trivial vacuum $\psi=0$ is suppressed in the large-$N$ limit. In addition, the large-$N$ limit favours supercooling; since the height of the potential barrier increases with $N$, the transition is delayed at the metastable trivial vacuum until it is on the verge of becoming unstable.

The model is a very simplified one, intended to demonstrate the effective action method for computing GWs from phase transitions in strongly coupled field theories. The observation that TeV-scale phase transitions lead to observable signals motivates the exploration of more realistic models.  The method also gives general predictions for the behaviour of the parameters with $N$.

The method does not yet allow us to compute $v_\text{w}$.  It would be very interesting to look for terms in the effective action coupling the scalar to the fluid, similar to those known to appear in weakly-coupled theories \cite{Ignatius:1993qn,Liu:1992tn,Moore:1995ua}.

%%%%%%%%%%%%%%%%%%%%%%%%%%%%
\begin{acknowledgments}
{\em  Acknowledgments} We thank Kari Rummukainen for useful discussions. F.~R.~A. has been supported through the STFC/UKRI grant no. 2131876. O.~H. has been supported by the Academy of Finland grant no.~1330346, the Ruth and Nils-Erik Stenb\"ack foundation, and the Waldemar von Frenckell foundation. M.~H. (ORCID ID 0000-0002-9307-437X) acknowledges support from the Academy of Finland grant no.~333609. C.~H. has been partially supported by the Spanish Ministerio de Ciencia, Innovaci\'on y Universidades through the grant PGC2018-096894-B-100. N.~J. has been supported by the Academy of Finland grant no.~1322307.
\end{acknowledgments}

\bibliographystyle{apsrev4-2}

\bibliography{biblio}

\end{document}